%% file: towner_CKM2010.tex
\documentclass[12pt]{article}
\usepackage{graphicx}

\newcommand\pubnumber{           }
\newcommand\pubdate{\today}

\def\texas{Cyclotron Institute, Texas A \& M University, \\
College Station, Texas 77843}
\def\support{\footnote{The work of JCH was supported by the US Department
of Energy under Grant DE-FG02-93ER40773 and by the Robert A Welch
Foundation under Grant A-1397.  IST would like to thank the
Cyclotron Institute of Texas A \& M University for its
hospitality during annual two-month visits.}}
\def\Title#1{\begin{center} {\Large #1 } \end{center}}
\def\Author#1{\begin{center}{ \sc #1} \end{center}}
\def\Address#1{\begin{center}{ \it #1} \end{center}}

\newcommand\pubblock{\rightline{\begin{tabular}{l} \pubnumber\\
         \pubdate  \end{tabular}}}
\newenvironment{Abstract}{\begin{quotation}  }{\end{quotation}}
\newenvironment{Presented}{\begin{quotation} \begin{center} 
             PRESENTED AT\end{center}\bigskip 
      \begin{center}\begin{large}}{\end{large}\end{center} \end{quotation}}

\input econfmacros.tex
\begin{document}
\begin{titlepage}
\pubblock

\vfill
\Title{$V_{ud}$ from Nuclear Decays} 
\vfill
\Author{ I.S. Towner and J.C. Hardy\support}
\Address{\texas}
\vfill
\begin{Abstract}
The current value of $V_{ud}$ is determined from
superallowed beta-decay experiments.  Other methods, briefly
summarised here, have to overcome specific experimental hurdles before
they are competitive.  However, the nuclear results do depend on a
nuclear-structure calculation of isospin-symmetry breaking, which is 
often a cause of some concern.  We show
here, by adopting the Conserved Vector Current (CVC) hypothesis, that
these theoretical corrections can be tested for consistency.   In this
test, calculations based on the shell model  
with Saxon-Woods radial
functions perform the best.

\end{Abstract}
\vfill
\begin{Presented}
$6^{th}$ International Workshop on the CKM Unitarity Triangle, \\
University of Warwick, UK, 6--10 September, 2010
\end{Presented}
\vfill
\end{titlepage}
\def\thefootnote{\fnsymbol{footnote}}
\setcounter{footnote}{0}

\section{Current value for $V_{ud}$ }

Superallowed $0^+ \rightarrow 0^+$ beta decay between isospin
$T = 1$ nuclear analog states currently provides the most
accurate determination of the Cabibbo-Kobayashi-Maskawa (CKM)
matrix element, $V_{ud}$.  There are two reasons for this:
First, there are many nuclear beta decays that could be 
chosen for study.  By limiting the study to just those decays
between $0^+$ analog states only the vector component of the
weak interaction is operative.  
Second, by limiting the
study in this way, the Conserved Vector Current (CVC) hypothesis
becomes useful.  This hypothesis states that the strength of the
vector component of the weak interaction, $G_V$, is a `true'
constant and independent of the nucleus under study.  This
result provides a consistency check among the different 
nuclear decays studied.  The hypothesis, however, is only
operative in the isospin-symmetry limit.  So one disadvantage
is that a nuclear-structure dependent calculation of 
isospin-symmetry breaking is required and the uncertainty associated 
with this is the subject of the second half of this report.
For the moment, we note that this correction is small and is
testable via the CVC hypothesis.

The analysis proceeds as follows.  The experimental measured
quantities of $Q$-value, lifetime and branching ratio are
combined into an $ft$ value.  To this, radiative and
isospin-symmetry breaking corrections are added in defining
a `corrected' ${\cal F}t$ value
\begin{equation}
{\cal F}t \equiv ft(1+\delta_R^{\prime})(1+\delta_{NS}-\delta_C) =
\frac{K}{2 G_V^2 (1+\Delta_R^V)} .
\label{Ftdef}
\end{equation}
Here the radiative correction is separated into three terms,
$\delta_R^{\prime}$, $\delta_{NS}$ and $\Delta_R^V$, where 
$\delta_R^{\prime}$ and $\delta_{NS}$ depend on the nucleus under
study, while $\Delta_R^V$ does not.  Further,
$\delta_R^{\prime}$ depends trivially on the nucleus, depending
on the total charge of the nucleus, $Z$, and on the emitted
electron's energy.  But $\delta_{NS}$, like the isospin-symmetry
breaking correction $\delta_C$ depends in its evaluation on the
details of nuclear structure.  Lastly, $K$ is a constant,
$K/(\hbar c)^6 = 2 \pi^3 \hbar \ln 2/(m_e c^2)^5$, with $m_e$
the electron mass.  Immediately one sees from the CVC
hypothesis that a nucleus-independent $G_V$ value leads to the
requirement that the ${\cal F}t$ value be nucleus-independent as well.
This provides a demanding consistency check on the
experimental measurements and the theoretical corrections.
If the ${\cal F}t$ values are found to be statistically consistent
with each other, then one is justified in taking an average
value, $\overline{{\cal F}t}$, from which $G_V$ can be determined.
In addition, via the relationship $V_{ud} = G_V / G_F$, where
$G_F$ is the well-known weak-interaction strength constant for a
purely leptonic decay, a value for $V_{ud}$ is obtained as well.

From the 2009 survey of experimental data, Hardy and Towner \cite{HT09}
determine
the value of $\overline{{\cal F}t}$ to be
\begin{equation}
{\cal F}t = 3071.81 \pm 0.83~s
\label{AvgFt}
\end{equation}
leading to
\begin{equation}
|V_{ud}| = 0.97425 \pm 0.00022 .~~~~~~~~~~[0 \rightarrow 0]
\label{Vud00}
\end{equation}
Other methods of obtaining $V_{ud}$ -- see survey in \cite{TH10} --
are currently less accurate.  They are:
\begin{itemize}
\item {\it neutron decay}, for which
\begin{equation}
|V_{ud}| = 0.9743 \pm 0.0015 .~~~~~~~~~~~~[{\rm neutron}]
\label{Vudneut}
\end{equation}
This value was presented at the CKM2010 Workshop by M\"{a}rkisch \cite{Ma10}.  It
is based on the 2010 Particle Data Group's analysis \cite{PDG10}, but
updated for new lifetime measurements from Serebrov {\it et al}. \cite{Se05}
and Pichlmaier {\it et al}. \cite{Pi10} and for preliminary beta-asymmetry
measurements from PERKEO II \cite{Ab08} and UCNA \cite{Li10}.
\item {\it $T = 1/2$ mirror transitions}, for which
\begin{equation}
|V_{ud}| = 0.9719 \pm 0.0017 ~~~~~~~~~~~~[{\rm mirror~transitions}]
\label{Vudmirror}
\end{equation}
from Naviliat-Cuncic and Severijns \cite{NS09}.
\item {\it pion beta decay}, for which
\begin{equation}
|V_{ud}| = 0.9742 \pm 0.0026 ~~~~~~~~~~~~[{\rm pion}]
\label{Vudpion}
\end{equation}
using the branching ratio measured by the PIBETA group \cite{Po04}.
\end{itemize}

The CKM matrix is posited to be unitary.  To date, the most demanding test
of this comes from the sum of squares of the top-row elements,
$|V_{ud}|^2+|V_{us}|^2+|V_{ub}|^2$,
which should sum to one.  Taking $|V_{us}|$ from the
recent FlaviaNet report \cite{An10}, $|V_{us}| = 0.2253(9)$,
and $|V_{ub}|$ from the Particle Data Group \cite{PDG10},
$|V_{ub}| = 3.39(44) \times 10^{-3}$, the unitarity sum becomes
\begin{equation}
|V_{ud}|^2+|V_{us}|^2+|V_{ub}|^2 = 0.99990 \pm 0.00060 .
\label{unit}
\end{equation}
This result shows unitarity to be fully satisfied to a precision
of $0.06 \%$.  Only $V_{us}$ and $V_{ud}$ contribute preceptibly
to the uncertainty and their contributions to the error budget
are almost equal to one another.

%%%%%%%%%%%%%%%%%%%%%%%%%%%%%%%%%%%%%%%%%%%%%%%%%%%%%%%%%%%%%%%%%%%%%%%%%
%%
%%   use this format to include an .eps figure into your paper
%%
%\begin{figure}[htb]
%\centering
%\includegraphics[height=1.5in]{magnet}
%\caption{Plan of the magnet used in the mesmeric studies.}
%\label{fig:magnet}
%\end{figure}
%%%%%%%%%%%%%%%%%%%%%%%%%%%%%%%%%%%%%%%%%%%%%%%%%%%%%%%%%%%%%%%%%%%%%%%%%%%

\section {Test of isospin-symmetry breaking correction}

Let's return to the isospin-symmetry breaking correction, $\delta_C$.
Its evaluation requires a nuclear-structure calculation.
Although the role played by nuclear structure is relatively small,
the precision currently reached by experiment is such that
the theoretical uncertainties introduced with $\delta_C$ now
dominate over the experimental uncertainties.  Consequently, this
correction has attracted a lot of attention recently.  We
offer a recommended set of values for this correction \cite{TH08,HT09},
but there are a growing number of alternative choices 
\cite{OB85,OB95,Sa96,Li09,Au09}.
There has also been a claim, albeit unsupported by any detailed
computations, that our calculations neglect a radial excitation
term, which is purported to be important \cite{MS08}.  To
counterbalance that, however, there are two recent papers that
confirm our result:  one \cite{Gr10} does so based on a semi-empirical
analysis of the data, while the other \cite{Sa10} quotes results
from a Skyrme-density-functional-theory calculation in which
simultaneous isospin and angular-momentum projection has been
incorporated.

Clearly it would be valuable if the various sets of $\delta_C$
corrections could be tested against the experimental data.
Towner and Hardy \cite{TH10a} have suggested such a test, which
is based on the acceptance of the CVC hypothesis.  We start
by rearranging Eq.~(\ref{Ftdef}) to read
\begin{equation}
\delta_C = 1 + \delta_{NS} - \frac{\overline{{\cal F}t}}{ft
(1 + \delta_R^{\prime})}
\label{dctest}
\end{equation}
where ${\cal F}t$ has been replaced by its average value.  For any
set of $\delta_C$ values to be acceptable, this equation must be satisfied.
For a series of $n$ superallowed transitions, one treats
$\overline{{\cal F}t}$ as a single adjustable parameter and use it
to bring the $n$ results from the right-hand side of Eq.~(\ref{dctest}),
which is based predominantly on the experimental $ft$ values
($\delta_{NS}$ is small, and $\delta_R^{\prime}$ unambiguous),
into the best possible agreement with the corresponding $n$
calculated values for $\delta_C$.  The normalized $\chi^2$, minimized 
by this process then provides a figure of merit for that set of
calculations.

The recent $\delta_C$ calculations are described in \cite{TH10a} and are:
\begin{itemize}
\item Shell model with Saxon-Woods radial functions, SM-SW \cite{TH08}.
\item Shell model with Hartree-Fock radial functions, SM-HF \cite{HT09}.
\item Relativistic Hartree-Fock with the random phase approximation (RPA)
and an effective interaction labelled PKO1, RHF-RPA \cite{Li09}.
\item Relativistic Hartree with RPA and a density-dependent effective
interaction, labelled DD-ME2, RH-RPA \cite{Li09}.
\item Isovector monopole resonance model, IVMR \cite{Au09}.
\end{itemize}
We have applied the test to these five sets of model calculations.
The resulting normalized $\chi^2$ for each least-squares fit --
expressed as $\chi^2/n_d$, where $n_d$ is the number of degrees of
freedom -- is given in Table~\ref{t:chi2}.

%%%%%%%%%%%%%%%%%%%%%%%%%%%%%%%%%%%%%%%%%%%%%%%%%%%%%%%%%%%%%%%%%%%%%%%%%
%%
%%   use this format to include a LaTeX table  into your paper
%%
\begin{table}[t]
\begin{center}
\begin{tabular}{lccccc}  
  & SM-SW & SM-HF & RHF-RPA & RH-RPA & IVMR \\ \hline
  & & & & & \\[-3mm]
$\chi^2/n_d$ (Row 1 -- see text) &
1.2 & 8.3 & 7.2 & 6.0 & 48.0 \\
Confidence Level (\%) &
26 & 0 & 0 & 0 & 0 \\
$\chi^2/n_d$ (Row 3 -- see text) &
0.4 & 2.2 & 2.7 & 2.1 & 11.0 \\
$\chi^2/n_d$ (Row 4 -- see text) &
0.3 & 1.1 & 1.6 & 1.3 & 4.5  \\[2mm] \hline
\end{tabular}
\caption{Normalized $\chi^2/n_d$ obtained in the test described in the
text for five sets of model calculations of $\delta_C$.
From Ref. \cite{TH10a}.}
\label{t:chi2}
\end{center}
\end{table}
%%%%%%%%%%%%%%%%%%%%%%%%%%%%%%%%%%%%%%%%%%%%%%%%%%%%%%%%%%%%%%%%%%%%%%%%%%%

We give three sets of normalized $\chi^2$; they differ one from another on
how the uncertainties are handled.  Strictly speaking, the $\chi^2$ test
only has an unambiguous interpretation if the errors considered are
solely statistical.  Thus in the first row in Table~\ref{t:chi2}, we keep
only statistical errors on the experimental $ft$ values and assign
no errors to the theoretical quantities $\delta_R^{\prime}$, $\delta_{NS}$
and $\delta_C$.  For this case, we can define a confidence level as
\cite{PDG10}
\begin{equation}
CL = \int_{\chi_0^2}^{\infty} P_{n_d}(\chi^2) d \chi^2
\label{CL}
\end{equation}
where $P_{n_d}(\chi^2)$ is the $\chi^2$ probability distribution
function for $n_d$ degrees of freedom, and $\chi_0^2$ is the minimum value
of $\chi^2$ obtained in the fit for the particular model set of
values for $\delta_C$.  Loosely speaking, the larger the value of $CL$
the more acceptable are the values of $\delta_C$ in satisfying
the CVC hypothesis.  The $CL$ values are given in the second row of
the Table.  In the third row, we have added non-statistical errors to
the radiative correction, while in the fourth row non-statistical
errors are included for both the radiative and isospin-symmetry
breaking corrections.  The inclusion of non-statistical errors
generally reduces the normalized $\chi^2$ of the fit, but the
ranking of the models remains unaltered.

The most obvious outcome of these analyses is that only one model,
SM-SW, produces satisfactory agreement with CVC.  All the others
have confidence levels below $0.5 \%$.  It is somewhat surprising
that SM-HF with Hartree-Fock radial functions does not do as well
as Saxon-Woods radial functions.  The problem is that these SM-HF 
calculations fail to give large enough $\delta_C$ values for high-$Z$
cases of $^{62}$Ga and $^{74}$Rb.  This has been noted before
by Ormand and Brown \cite{OB95}.  We have tried varying the
Skyrme interaction used in the Hartree-Fock calculation -- to date
we have sampled 12 interactions -- but they all fail in the high-$Z$
cases.  The discrepancy appears inherent in the SM-HF model.

\end{document}

%% file: econfmacros.tex
\def\beq{\begin{equation}}
\def\eeq#1{\label{#1}\end{equation}}
\def\eeqn{\end{equation}}

\def\beqa{\begin{eqnarray}}
\def\eeqa#1{\label{#1}\end{eqnarray}}
\def\eeqan{\end{eqnarray}}

\let\bar=\overbar

\def\Dslash{\not{\hbox{\kern-4pt $D$}}}
\def\dslash{\not{\hbox{\kern-2pt $\del$}}}

\def\msb{{\bar{\ssstyle M \kern -1pt S}}}